\documentclass[12pt]{article}

\usepackage{mathptmx,mathrsfs}
\usepackage{amsfonts,amsmath,amssymb,amsthm}
\usepackage{indentfirst}
\usepackage{cite}
\usepackage{hyperref}

\numberwithin{equation}{section}
\begin{document}

\title{Higher-derivative non-Abelian gauge fields via the Faddeev-Jackiw
formalism}
\author{R. Bufalo$^{1,2}$\thanks{%
rbufalo@ift.unesp.br} ~ and B.M. Pimentel$^{2}$\thanks{%
pimentel@ift.unesp.br} \\
\textit{{$^{1}${\small Department of Physics, University of Helsinki, P.O. Box 64}}}\\
\textit{\small FI-00014 Helsinki, Finland}\\
\textit{{$^{2}${\small Instituto de F\'{\i}sica Te\'orica (IFT), UNESP,  S\~ao Paulo State University}}} \\
\textit{\small Rua Dr. Bento Teobaldo Ferraz 271, Bloco II Barra Funda, CEP
01140-070 S\~ao Paulo, SP, Brazil}\\
}
\maketitle
\date{}

\begin{abstract}
In this paper we analyze two higher-derivative theories, the generalized
electrodynamics and the Alekseev-Arbuzov-Baikov's effective Lagrangian from
the point of view of Faddeev-Jackiw sympletic approach. It is shown that the
full set of constraint is obtained directly from the zero-mode eigenvectors,
and that they are in accordance with known results from Dirac's theory, a
remnant and recurrent issue in the literature. The method shows to be rather
economical in relation to the Dirac's one, obviating thus unnecessary
classification and calculations. Afterwards, to conclude we construct the
transition-amplitude of the non-Abelian theory following a constrained
BRST-method.
\end{abstract}

\newpage


\section{Introduction}
\label{intro}

A standard classical treatment of constrained theories was given originally
by Dirac \cite{25}, it essentially analyzes the canonical structure of any
theory, and it has been widely used in a great variety of quantum systems.
However, it could be realized that Dirac's methodology is unnecessarily
cumbersome and can be streamlined. Within this context, Faddeev and Jackiw
\cite{1} suggested a sympletic approach for constrained systems based a
first-order Lagrangian. This method has some very interesting properties of
obviating the constraint classification, unnecessary calculations and the
hypothesis of Dirac's conjecture as well. The Faddeev-Jackiw (FJ) sympletic
formalism has been studied in a systematic way in different scenarios,
shedding a new light into the research of constrained dynamics.

The basic geometric structure of the Faddeev-Jackiw theory can be read
directly from the elements of the inverse sympletic matrix, and coincides
with the correspondent Dirac brackets, providing thus a bridge to the
commutators of the quantized theory. On the other hand, the results obtained
from the Faddeev-Jackiw approach have been compared with the corresponding
results of Dirac method in different situations, for unconstrained and
constrained systems, but it is still matter of study.

The method has as the key ingredient that these constraints produce
deformations in the two-form sympletic matrix in such way that, when all
constraints are considered (by means of a Darboux transformation), the
sympletic matrix is non-singular. As a result, it was obtained the Dirac
brackets. Nevertheless, it is important to emphasize that sometimes, it
happens that the (iteratively deformed \footnote{%
Actually the geometric role played by the constraints is to produce a
'deformation' in the original, singular, sympletic two-form matrix.})
two-form matrix is singular and no new constraint is obtained from the
corresponding zero-mode. This is the case when one deals with gauge
theories. At this point one should introduce convenient gauge (subsidiary)
conditions like a constraint and the two-form matrix becomes, therefore,
invertible. This extension was proposed and developed by Barcelos-Neto and
Wotzasek \cite{2}, and studied in several models \cite{9}. It basically
followed the spirit of Dirac's work, with proposal works by imposing the
stability of the constraints under time evolution. So, constraints are not
solved but embedded in an extended phase-space. This is a more suitable
approach when some relevant symmetries must be preserved.

A subtle issue subsequent to the Faddeev-Jackiw method is its equivalence to
the Dirac method. Initially the equivalence was discussed in cases when the
systems have not constraint \cite{16}; but, in a constrained system, the
situation was not completely clear, and some argumentation was provided
earlier \cite{4} about the equivalence between the methods. However,
recently it was presented a proof \cite{5} that the usual Faddeev-Jackiw
method and Dirac method were not completely equivalent; namely, they showed
that some constraints calculated in Dirac formalism do not appear in the
calculation in Faddeev-Jackiw formalism. And then these would result in the
contradiction between the usual Faddeev- Jackiw quantization and Dirac
quantization \cite{3}.

Higher-derivative Lagrangian functions \cite{14} are a fairly interesting
branch of the ongoing effective theories, and were initially proposed as an
attempt to enhance and render a better ultraviolet behavior of physically
relevant models. It is known that higher-derivative theories have, as a field theory,
better renormalization properties than the conventional ones. These
properties have shown to be quite appealing in the attempts to have a
quantized and renormalizable theory of gravity \cite{18}. The undesired
features of the higher-derivative theory is that they possess a Hamiltonian
that is not bounded from below and that the process of adding such terms
jeopardizes the unitarity of the theory \cite{19}. Besides all these
motivations we emphasize that, from a theoretical point of view,
higher-derivative theories have many interesting features that justify their
study by itself.

As it has been pointed out in several works \cite{17,12,21} along the years,
it is long clear that Maxwell's theory is not the only one to describe the
electromagnetic field. One of the most successful generalizations is the
generalized electrodynamics \cite{17}. Actually, Podolsky's theory is the
only one linear, Lorentz, and $U(1)$ invariant generalization of Maxwell's
theory \cite{21}. Another interesting feature inherent to Podolsky's theory
is the existence of a generalized gauge condition also, namely, the
generalized Lorenz condition: $\Omega[A]=\left(1+M^{-2}\square\right)%
\partial_\mu A^{\mu}$; considered an important issue, it is only through the
choice of the correct gauge condition that we can completely fix the gauge
degrees of freedom of a given gauge theory \cite{12}. The relative success
of these achievements motivated some authors to propose finite extensions of
Quantum Chromodynamics (QCD) \cite{10} and also to advocate that higher
order terms would be able to explain the quark confinement. Our main goal
here would be exactly to study both higher-derivative theories, Podolsky's
electrodynamics and a non-Abelian \cite{15} extension of the model, also
known as the Alekseev-Arbuzov-Baikov's effective Lagrangian \cite{11} in the
framework Faddeev-Jackiw sympletic approach. As far we have no knowledge of
application of the Faddeev-Jackiw method to higher-derivative theories.
Moreover, it may also shed some new light on the issue of whether the
accordance between the Dirac and Faddeev-Jackiw methods holds.

In this paper, we discuss the canonical structure of the Podolsky's
electrodynamics and the $SU(N)$ Alekseev-Arbuzov-Baikov's effective
Lagrangian in the light of the Faddeev-Jackiw approach. In Sect.\ref{sec:1}
we start by making a brief review of both the FJ and (constrained) sympletic
formalisms. And, as the generalized electrodynamics of Podolsky has been
already subject of analysis from the Dirac's point of view \cite{12}, we shall study
the theory via the FJ method in order to present an exercise of the
methodology and also to check its consistency. Next, in Sect.\ref{sec:2}, we
discuss and introduce the Alekseev-Arbuzov-Baikov's effective Lagrangian by
discussing the generalized electrodynamics by making use of enlargement of
the gauge group to non-Abelian ones. Having defined the Lagrangian density
we proceed in presenting the methodology, and obtaining the full set of
constraints of the theory. Although an attempt of a path-integral
formulation based on the FJ method has been proposed \cite{8}, there is no
conclusive, neither clear argument to show the consistence of the method.
Therefore, by means of complementarity of the previous discussion, we
conclude the section by constructing the transition-amplitude for the
non-Abelian theory via the Batalin-Fradkin-Vilkovisky (BFV) method \cite{22}%
, obtaining an important outcome for subsequent analysis in the quantum
level. In Sect.\ref{sec:6} we summarize the results, and present our final
remarks and prospects.
\section{Generalized electrodynamics via Faddeev-Jackiw formalism}

\label{sec:1}

\subsection{Faddeev-Jackiw sympletic method}

Let us start with a first-order in time derivative Lagrangian, which may
arise from a conventional second-order one after introducing auxiliary
fields. First, one can construct the sympletic Lagrangian \footnote{%
In this section we discuss a system with finite degree of freedom. However,
an extension to the infinite degree of freedom case can be attained in a
straightforward way.}%
\begin{equation}
\mathcal{L}=a_{i}\left( \xi \right) \dot{\xi}^{i}-\mathcal{V}\left( \xi
\right) ,  \label{eq 1.0}
\end{equation}%
with the arbitrary one-form components $a_{i}$, with $i=1,...,N$. The
first-order system is characterized by a closed two-form. If the two-form is
non-degenerated, it defines a sympletic structure on the phase space,
described by the coordinates $\xi _{i}$. On the other hand, if the two-form
is singular, with constant rank, it is called a pre-sympletic two-form.
Thus, in terms of components, the (pre)sympletic form is defined by%
\begin{equation}
f_{ij}=\frac{\partial }{\partial \xi ^{i}}a_{j}\left( \xi \right) -\frac{%
\partial }{\partial \xi ^{j}}a_{i}\left( \xi \right) .  \label{eq 1.1}
\end{equation}%
The Euler-Lagrange equations are given by%
\begin{equation}
f_{ij}\dot{\xi}^{j}=\frac{\partial }{\partial \xi ^{i}}\mathcal{V}\left( \xi
\right) .  \label{eq 1.2}
\end{equation}%
Now, when the two-form $f_{ij}$ is nonsingular, it has an inverse $f^{ij}$,
then%
\begin{equation}
\dot{\xi}^{i}=f^{ij}\frac{\partial }{\partial \xi ^{j}}\mathcal{V}\left( \xi
\right) ,  \label{eq 1.3}
\end{equation}%
and the basic bracket is defined as $\left\{ \xi ^{i},\xi ^{j}\right\} =f^{ij} $.
However, in the case that the Lagrangian \eqref{eq 1.0} describes a
constrained system, the sympletic matrix is singular, and the constraints
hidden in the system need to be determined. Let us suppose that the rank of $%
f_{ij}$ is $2n$, so there is $N-2n=M$ zero-mode vectors $\mathbf{v}^{\alpha
} $, $\alpha =1,...,M$. The system is then constrained by $M$ equation in
which no time-derivatives appear. Then there will be constraints that reduce
the number of degrees of freedom. Thus, multiplying \eqref{eq
1.2} by the (left) zero-modes $\mathbf{v}^{\alpha }$ of $f_{ij}$ we get the
(sympletic) constraints in the form of algebraic relations%
\begin{equation}
\Omega ^{\alpha }\equiv \mathbf{v}_{i}^{\alpha }\frac{\partial }{\partial
\xi ^{i}}\mathcal{V}\left( \xi \right) =0.  \label{eq 1.5}
\end{equation}%
Then, one can give the first-iterated Lagrangian by introducing
corresponding Lagrange multipliers of the obtained constraints%
\begin{equation}
\mathcal{L}=a_{i}^{\left( 1\right) }\left( \xi \right) \dot{\xi}^{i}+\Omega
^{\alpha }\lambda _{\alpha }-\mathcal{V}^{\left( 1\right) }\left( \xi
\right) .
\end{equation}%
Hence, one may regard the introduced Lagrange multipliers $\lambda $ as
sympletic variables and can extend the sympletic variable set. This
procedure reduces the number of $\xi $'s. Then the whole procedure can be
repeated again until all constraints are eliminated and we are left with a
completely reduced, unconstrained, and canonical system. However, it should
be remarked that in the case of gauge theories, the zero-mode does not give
any new constraint (it still does not give the full rank matrix), and the
sympletic matrix remains singular. Thus, we should consider that it is
necessary to introduce gauge condition(s) to obviate the singularity. So the
work can be finished in expectation in terms of the original variables, and
the basic brackets can be determined.

\subsection{Generalized electrodynamics}

The purpose of the present study is to examine the FJ methodology applied in
the analysis of a higher-derivative theory. It is interesting to study
first, as a simpler example, an Abelian theory. Therefore, in order to keep
the things in a simple realm, we choose the simplest but rather interesting
Abelian electrodynamics of Podolsky, whose Lagrangian density is given by%
\begin{equation}
\mathcal{L}=-\frac{1}{4}F^{\mu \nu }F_{\mu \nu }+\frac{1}{2M^{2}}\partial
_{\mu }F^{\mu \nu }\partial ^{\lambda }F_{\lambda \nu },  \label{eq 0.0}
\end{equation}%
where $F_{\mu \nu }=\partial _{\mu }A_{\nu }-\partial _{\nu }A_{\mu }$ and
the spacetime metric elements are $\eta _{\mu \nu }=\left( 1,-1,-1,-1\right)
$. It should be mentioned that we shall follow the Ostrogradski approach
\cite{14} to deal with higher-derivative terms. Hence, it should be
introduced another set of canonical pair $\left( \Gamma ^{\mu }\equiv
\partial _{0}A^{\mu },\phi _{\nu }\right) $ in order to have a correct
expanded phase space to thus proceed with the canonical analysis. With this
thought in mind one finds then the following Lagrangian \cite{12}
\begin{align}
\mathcal{L}=& \frac{1}{2}\left( \overrightarrow{\Gamma }-\nabla A_{0}\right)
^{2}+\frac{1}{2}\left( \nabla \times \overrightarrow{A}\right) ^{2}\notag \\
 &+\frac{1}{2M^{2}}\bigg[ \left( \nabla .\overrightarrow{\Gamma }%
-\nabla ^{2}A_{0}\right) ^{2}-\left( \partial _{0}\overrightarrow{\Gamma }%
-\nabla \Gamma _{0}-\nabla \times \left( \nabla \times \overrightarrow{A}%
\right) \right) ^{2}\bigg] . \label{eq 0.5}
\end{align}%
To transform this Lagrangian from second to first-order, we shall use an
auxiliary field, that is choose to be the canonical momentum due to an
algebraic simplification that it provides. In that case, we should remind
that we have additional set of canonical pairs, in particular here, $\left(
A,\pi \right) $ and $\left( \Gamma ,\phi \right) $:%
\begin{equation}
\phi ^{\mu }=\frac{\partial \mathcal{L}}{\partial \left( \partial
_{0}\partial _{0}A_{\mu }\right) },  \label{eq 2.5}
\end{equation}%
and%
\begin{equation}
\pi ^{\mu }=\frac{\partial \mathcal{L}}{\partial \left( \partial _{0}A_{\mu
}\right) }-2\partial _{k}\left( \frac{\partial \mathcal{L}}{\partial
\left( \partial _{0}\partial _{k}A_{\mu }\right) }\right) -\partial
_{0}\left( \frac{\partial \mathcal{L}}{\partial \left( \partial _{0}\partial
_{0}A_{\mu }\right) }\right) ,  \label{eq 2.6}
\end{equation}%
resulting into the following expressions%
\begin{equation}%
\pi ^{\mu }= F^{\mu 0}-M^{-2}\left( \eta ^{\mu k}\partial _{k}\partial
_{\lambda }F^{0\lambda }-\partial _{0}\partial _{\lambda }F^{\mu \lambda
}\right) ,
\end{equation}
and
\begin{equation}%
M^{2}\phi ^{\mu } =\left( \eta ^{\mu 0}\partial _{\lambda }F^{0\lambda
}-\partial _{\lambda }F^{\mu \lambda }\right) ,  \notag
\end{equation}%
where%
\begin{equation}
M^{2}\vec{\phi}=\partial _{0}\overrightarrow{\Gamma }-\nabla \Gamma
^{0}-\left( \nabla \times \left( \nabla \times \overrightarrow{A}\right)
\right) .  \label{eq 0.6}
\end{equation}%
In order to obtain the quadratic kinetic terms we may make use of the
equation of motion for $\phi $ \eqref{eq 0.6} back into the Lagrangian.
Therefore, in this case one can cast the Lagrangian density \eqref{eq 0.5} as%
\begin{equation}
\mathcal{L}=-\phi _{k}\dot{\Gamma}_{k}+\pi _{\mu }\dot{A}^{\mu }-\mathcal{V}%
^{\left( 0\right) },  \label{eq 0.8}
\end{equation}%
where the potential density is%
\begin{align}
\mathcal{V}^{\left( 0\right) }=& \pi _{\mu }\Gamma ^{\mu }-\frac{1}{2}\left(
\overrightarrow{\Gamma }-\nabla A_{0}\right) ^{2}-\frac{1}{2}\left( \nabla
\times \overrightarrow{A}\right) ^{2}-\frac{1}{2M^{2}}\left( \nabla .%
\overrightarrow{\Gamma }-\nabla ^{2}A_{0}\right) ^{2}  \notag \\
& -\frac{M^{2}}{2}\overrightarrow{\phi }^{2}-\overrightarrow{\phi }%
.\left( \nabla \Gamma _{0}+\nabla \times \left( \nabla \times
\overrightarrow{A}\right) \right) .  \label{eq 0.9}
\end{align}%
The initial set of sympletic variables is seen to be $\xi _{\alpha }^{\left( 0\right) }=\left\{ A_{k},\pi _{k},A_{0},\pi
_{0},\Gamma _{k},\phi _{k},\Gamma _{0}\right\}$, this permits us to identify the non-null canonical one-form%
\begin{equation}
^{\Gamma }a_{i}^{\left( 0\right) }=-\phi _{i},\quad ^{A}a_{i}^{\left(
0\right) }=-\pi _{i},\quad ^{A_{0}}a^{\left( 0\right) }=\pi _{0}.
\label{eq 0.11}
\end{equation}%
These previous results lead to the corresponding two-form matrix%
\begin{equation}
^{\left( 0\right) }f_{ab}\left( x,y\right) =\left[
\begin{array}{cc}
\mathbf{A}_{ij}  ~&~ \mathbf{0}_{\mathbf{4\times 3}} \\
\mathbf{0}_{\mathbf{3\times 4}} ~& ~\mathbf{B}_{ij}
\end{array}%
\right]\delta \left( x,y\right) ,
\end{equation}%
with%
\begin{equation}
\mathbf{A}_{ij}=\left[
\begin{array}{cccc}
0 & \delta _{ij} & 0 & 0 \\
-\delta _{ij} & 0 & 0 & 0 \\
0 & 0 & 0 & -1 \\
0 & 0 & 1 & 0%
\end{array}%
\right]  ,\quad \mathbf{B}_{ij} =\left[
\begin{array}{ccc}
0 & \delta _{ij} & 0 \\
-\delta _{ij} & 0 & 0 \\
0 & 0 & 0%
\end{array}%
\right] ,  \label{eq 0.12}
\end{equation}%
it is not complicated to see that the matrix is singular. Moreover, it is
easy to determine that the eigenvector with zero eigenvalue is%
\begin{equation}
\nu _{\alpha }=\left( \mathbf{0},\mathbf{0},0,0,\mathbf{0},\mathbf{0},\nu ^{%
\mathbf{7}}\right) ,  \label{eq 0.14}
\end{equation}%
where $\nu ^{7}$ is arbitrary and associated to $\Gamma _{0}$. Therefore,
from the eigenvector $\nu _{\alpha }$ \eqref{eq 0.14} we can evaluate the
consistence condition as%
\begin{equation}
\int dx dy\nu ^{7} \frac{\delta }{\delta \Gamma _{0}\left(
x\right) } \mathcal{V}^{\left( 0\right) }\left( y\right) = \int dx\nu ^{7} \left( \pi _{0}+\nabla .\overrightarrow{\phi }\right) =0,
\label{eq 0.15}
\end{equation}%
since $\nu ^{7}\left( x\right) $ is an arbitrary function, we obtain the
constraint%
\begin{equation}
\Omega \left( x\right) \equiv \pi _{0}\left( x\right) +\nabla .%
\overrightarrow{\phi }\left( x\right) =0.  \label{eq 0.16}
\end{equation}%
Introducing this constraint back into the Lagrangian by means of a Lagrange
multiplier $\lambda $ \footnote{%
It should be noted that when the constraint $\Omega $ is imposed the
dependence in $\Gamma _{0}$ naturally disappears.}%
\begin{equation}
\mathcal{L}=-\phi _{k}\dot{\Gamma}_{k}+\pi _{\mu }\dot{A}^{\mu }+\dot{\lambda%
}\left( \pi _{0}+\nabla .\overrightarrow{\phi }\right) -\mathcal{V}^{\left(
1\right) },  \label{eq 0.17}
\end{equation}%
where the first-iterated potential density is $\mathcal{V}^{\left( 1\right) }= \left. \mathcal{V}^{\left( 0\right)
}\right\vert _{\Omega =0} $ with
\begin{align}
\mathcal{V}^{\left( 1\right) }& =-\pi _{k}\Gamma _{k}-\frac{1}{2}\left( \overrightarrow{\Gamma }-\nabla
A_{0}\right) ^{2}-\frac{1}{2}\left( \nabla \times \overrightarrow{A}\right)
^{2}  -\frac{1}{2M^{2}}\left( \nabla .\overrightarrow{\Gamma }-\nabla
^{2}A_{0}\right) ^{2}\notag \\
&-\frac{M^{2}}{2}\overrightarrow{\phi }^{2}-%
\overrightarrow{\phi }.\left( \nabla \times \left( \nabla \times
\overrightarrow{A}\right) \right) .  \label{eq 0.18}
\end{align}%
From the above Lagrangian we have the following vectors%
\begin{gather}
^{\Gamma }a_{i}^{\left( 1\right) }=-\phi _{i}, \quad ^{A}a_{i}^{\left(
1\right) }=-\pi _{i}, \quad ^{A_{0}}a^{\left( 1\right) } =\pi _{0}, \quad ^{\lambda }a^{\left( 1\right) }=\pi _{0}+\nabla .\overrightarrow{\phi },
\label{eq 0.19}
\end{gather}%
these results lead to the corresponding two-form matrix%
\begin{equation}
^{\left( 1\right) }f_{ab}\left( x,y\right) =\left[
\begin{array}{cc}
\mathbf{A} _{ij}~ &~ \mathbf{D}_{j}  \\
\mathbf{-D}^{\mathbf{T}}_{i} ~&~ \mathbf{C}_{ij}
\end{array}%
\right] \delta \left( x,y\right),  \label{eq 0.20}
\end{equation}%
with $\mathbf{C}\left( x,y\right) $ and $\mathbf{D}\left( x,y\right) $ being
the Abelian version of the non-Abelian expressions $\mathbf{C}_{ab}\left(
x,y\right) $ and $\mathbf{D}_{ab}\left( x,y\right) $, Eq.\eqref{eq 2.24}. We
obtain once again a singular matrix. From that, we can determine its
eigenvector with zero eigenvalue,%
\begin{equation}
\overline{\nu }_{\alpha }=\left( \mathbf{0},\mathbf{0},\overline{\nu }^{3},0,%
\overline{\nu }_{i}^{5},\mathbf{0},\overline{\nu }^{\mathbf{7}}\right) ,
\label{eq 0.21}
\end{equation}%
Therefore, following the routine, from this eigenvector $\overline{\nu }%
_{\alpha }$ \eqref{eq 0.21} we can evaluate the consistence condition%
\begin{align}
\int dx\left[ \overline{\nu }^{3} \frac{\delta }{\delta
A_{0}\left( x\right) }+\overline{\nu }_{i}^{5}\frac{\delta }{%
\delta \Gamma ^{i}\left( x\right) }\right] \int dy\mathcal{V}^{\left(
1\right) }\left( y\right)= \int dx\overline{\nu }^{3}\left( \nabla .\overrightarrow{\pi }\right)
\left( x\right) =0,
\end{align}%
where in the last equality we have made use of the relation $\overline{\nu }%
_{i}^{5}-\partial _{i}\overline{\nu }^{3}=0$. Once again, as $\overline{\nu }%
^{3}$ is an arbitrary function, we obtain a new constraint relation (Gauss'
law)%
\begin{equation}
\bar{\Omega}\left( x\right) \equiv \left( \nabla .\overrightarrow{\pi }%
\right) \left( x\right) =0.  \label{eq 0.23}
\end{equation}%
Now, following the methodology, the second-iterated Lagrangian reads%
\begin{equation}
\mathcal{L}=-\phi _{k}\dot{\Gamma}_{k}+\pi _{\mu }\dot{A}^{\mu }+\dot{\lambda%
}\left( \pi _{0}+\nabla .\overrightarrow{\phi }\right) +\dot{\eta}\left(
\nabla .\overrightarrow{\pi }\right) -\mathcal{V}^{\left( 2\right) },
\label{eq 0.24}
\end{equation}%
whereas the second-iterated potential density is%
\begin{equation}
\mathcal{V}^{\left( 2\right) }=\left. \mathcal{V}^{\left( 1\right)
}\right\vert _{\bar{\Omega}=0}=\mathcal{V}^{\left( 1\right) }.
\label{eq 0.25}
\end{equation}%
From the above Lagrangian one finds the following vectors%
\begin{gather}
^{\Gamma }a_{i}^{\left( 2\right) }=-\phi _{i},\quad ^{A}a_{i}^{\left(
2\right) }=-\pi _{i},\quad ^{A_{0}}a^{\left( 2\right) }=\pi _{0}, \quad
^{\lambda }a^{\left( 2\right) }=\pi _{0}+\nabla .\overrightarrow{\phi }%
,\quad ^{\eta }a^{\left( 2\right) }=\nabla .\overrightarrow{\pi },
\end{gather}%
these results lead to the corresponding two-form matrix%
\begin{equation}
^{\left( 2\right) }f_{ab}\left( x,y\right) =\left[
\begin{array}{cc}
\mathbf{A}_{ij}  ~&~ \mathbf{E}_{j,x} \\
\mathbf{-E}_{i,y}^{T}~ &~ \mathbf{F}_{ij}
\end{array}%
\right] \delta \left( x,y\right) ,
\end{equation}%
with%
\begin{equation}
\mathbf{E}_{i,x}=\left[
\begin{array}{cccc}
0 & 0 & 0 & 0 \\
0 & 0 & 0 & -\partial _{i}^{x} \\
0 & 0 & 0 & 0 \\
0 & 0 & 1 & 0%
\end{array}%
\right]  ,\quad \mathbf{F}_{ij} =\left[
\begin{array}{cccc}
0 & \delta _{ij} & 0 & 0 \\
-\delta _{ij} & 0 & -\partial _{i}^{x} & 0 \\
0 & -\partial _{i}^{x} & 0 & 0 \\
0 & 0 & 0 & 0%
\end{array}%
\right] ,  \label{eq 0.26}
\end{equation}%
It follows that the second-iterated matrix $^{\left( 2\right) }f_{ab}\left(
x,y\right) $ is also singular. From that one obtain two zero-mode vectors%
\begin{equation}
\tilde{\nu}_{\alpha }=\left( \tilde{\nu}_{i}^{1},\mathbf{0},0,0,\mathbf{0},%
\mathbf{0},0,\tilde{\nu}^{\mathbf{8}}\right) ,
\end{equation}%
and%
\begin{equation}
\overline{\overline{\nu }}_{\alpha }=\left( \mathbf{0},\mathbf{0},\overline{%
\overline{\nu }}^{3},0,\overline{\overline{\nu }}_{i}^{5},\mathbf{0},%
\overline{\overline{\nu }}^{\mathbf{7}},\mathbf{0}\right) ,
\end{equation}%
However, the vector $\overline{\overline{\nu }}_{\alpha }$ generates the
constraint $\left( \nabla .\overrightarrow{\pi }\right) =0$. Therefore, it
is only the vector $\tilde{\nu}_{\alpha }$ of interest. Subsequently, the
consistence condition results into%
\begin{equation}
\int dxdy\left[ \tilde{\nu}_{i}^{1}\left( x\right) \frac{\delta }{\delta
A^{i}\left( x\right) }+\tilde{\nu}^{8}\left( x\right) \frac{\delta }{\delta
\eta \left( x\right) }\right] \mathcal{V}^{\left( 2\right) }\left( y\right)
=0,
\end{equation}%
Thus, the zero-mode does not generate any new constraints and, consequently,
the sympletic matrix remains singular. Being this a imprint characteristic
of gauge theories, therefore, the gauge degrees of freedom has to be fixed.
We choose the work here with the generalized Coulomb gauge: $%
A_{0}=0$ and $\left(1+M^{-2}\square \right) \left( \nabla .\overrightarrow{A}\right) =0$. \footnote{%
It is worth to mention that the complete generalized Coulomb gauge have in addition
the condition: $\Gamma _{0}=0$, but as it has already disappeared in the
Lagrangian, it is not necessary to impose it.} We then obtain a new
Lagrangian density%
\begin{align}
\mathcal{L}=-\phi _{k}\dot{\Gamma}_{k}+\pi _{k}\dot{A}^{k}+\dot{\lambda}%
\left( \pi _{0}+\nabla .\overrightarrow{\phi }\right) +\dot{\eta}\left(
\nabla .\overrightarrow{\pi }\right) +\dot{\chi}\left( 1-M^{-2}\nabla
^{2}\right) \nabla .\overrightarrow{A}-\mathcal{V}^{\left( 3\right) },
\label{eq 0.27}
\end{align}%
where the third-iterated potential density is $\mathcal{V}^{\left( 3\right) }= \left. \mathcal{V}^{\left( 2\right)
}\right\vert _{\bar{\Omega}=0}$ with
\begin{align}
\mathcal{V}^{\left( 3\right) }= -\pi _{k}\Gamma _{k}-\frac{1}{2}\overrightarrow{\Gamma }^{2}+\frac{1}{2}%
\overrightarrow{A}.\left( \nabla ^{2}\overrightarrow{A}\right) -\frac{1}{%
2M^{2}}\left( \nabla .\overrightarrow{\Gamma }\right) ^{2}-\frac{M^{2}}{2}%
\overrightarrow{\phi }^{2}-\overrightarrow{\phi }.\left( \nabla ^{2}%
\overrightarrow{A}\right) .  \label{eq 0.28}
\end{align}%
and we have absorbed the $\left( \nabla .\overrightarrow{A}\right) $ terms
into the new constraint. It is worth to emphasize that from the expression
for the potential $\mathcal{V}^{\left( 3\right) }$ one may reads which are
the dynamical variables; for instance, here, it consists into the canonical
set $\left\{ A_{k},\pi ^{m}\right\} $ and $\left\{ \Gamma _{k},\phi
^{m}\right\} $.

Nevertheless, from the above Lagrangian follows the vectors%
\begin{gather}
^{\Gamma }a_{i}^{\left( 3\right) }=-\phi _{i}, \quad ^{A}a_{i}^{\left(
3\right) }=-\pi _{i}, \quad ^{\lambda }a^{\left( 3\right) }=\pi _{0}+\nabla .%
\overrightarrow{\phi }, \notag \\
 ^{\eta }a^{\left( 3\right) }=\nabla . \overrightarrow{\pi },\quad ^{\chi }a^{\left( 3\right) }=\nabla ^{2}_{P}
\nabla .\overrightarrow{A},
\end{gather}%
where $\nabla ^{2}_{P}=\left( 1-M^{-2}\nabla ^{2}\right)$, these lead to the
corresponding third-iterated sympletic matrix \footnote{$\tilde{\mathbf{F}}_{ij,x}$ is equal to the expression
of $\mathbf{F}_{ij,x}$ Eq.\eqref{eq 0.26}, but with an additional fifth null line and column.}
\begin{equation}
^{\left( 3\right) }f_{ab}\left( x,y\right) =\left[
\begin{array}{cc}
\mathbf{B}_{ij} ~&~ \mathbf{G}_{j,x} \\
-\mathbf{G}_{i,y}^{T}~ &~ \tilde{\mathbf{F}}_{ij,x}
\end{array}%
\right] \delta \left( x,y\right),
\end{equation}%
with%
\begin{equation}
\mathbf{G}_{i,x}=\left[
\begin{array}{ccccc}
0 & 0 & 0 & 0 & -\nabla _{P}^{2}\partial _{i}^{x} \\
0 & 0 & 0 & -\partial _{i}^{x} & 0 \\
0 & 0 & 1 & 0 & 0%
\end{array}%
\right].
\end{equation}
This $^{\left( 3\right) }f_{ab}\left( x,y\right) $ is clearly a nonsingular
matrix and the corresponding inverse is easily obtained by a simple, but
rather lengthy calculation. Moreover, we may relabel $\lambda =\phi _{0}$, $%
\eta =\Gamma _{0}$, and $\chi =A_{0}$. Therefore, the generalized brackets
between the dynamical variables, the corresponding Dirac brackets in the
generalized radiation gauge, are just the elements of the inverse of such a
matrix, and reads%
\begin{align}
\left\{ A_{k}\left( x\right) ,\pi ^{m}\left( y\right) \right\} ^{\star }
&=\delta _{k}^{m}\delta \left( x,y\right) -\nabla ^{2}_{P} \partial
_{k}\partial ^{m}G\left( x,y\right) ,  \label{eq 0.33} \\
\left\{ \Gamma _{k}\left( x\right) ,\phi ^{m}\left( y\right) \right\}
^{\star } &=\delta _{k}^{m}\delta \left( x,y\right) .  \label{eq 0.33a}
\end{align}%
where we have introduced the Green's function
\begin{equation}
\nabla ^{2}_{P} \nabla ^{2}G\left( x,y\right) =\delta ^{\left( 3\right)
}\left( x,y\right) .
\end{equation}%
These results are in accordance to those obtained previously through an
analysis \textit{\`{a} la} Dirac in \cite{12}. Though we have obtained the
correct brackets to the dynamic variables, we are left with the whole
canonical variables (including the kinematical ones) without any trace of
which variables are in fact dynamical and that, therefore, should be
submitted to the quantization (a natural outcome of the Dirac's theory).
Nevertheless, the analysis of this particular theory showed to us that the
outcome of both theories match, although both present pros and cons,
especially those involving unnecessary calculation and tedious algebraic work.


\section{$SU\left(N\right) $ higher-derivative Yang-Mills-Utiyama theory}

\label{sec:2}

In this section we will go a step further from the previous discussion, and
consider an non-Abelian extension of the Podolsky's theory, also known as
the Alekseev-Arbuzov-Baikov's effective Lagrangian \cite{11}. This theory
was originally proposed to eliminate infrared divergences in $SU\left(
N\right) $ theories \cite{10}. In order to introduce some concepts, let us
consider the $U\left( 1\right) $ electrodynamics in four dimensions \eqref{eq 0.0}%
\begin{equation}
\mathcal{L}=-\frac{1}{4}F_{\mu \nu }F^{\mu \nu }+\frac{1}{2M^{2}}\partial
_{\mu }F^{\mu \nu }\partial ^{\lambda }F_{\lambda \nu }.  \label{eq 2.0}
\end{equation}%
Moreover, to make contact to the non-Abelian theory \cite{15}, it is
interesting to discuss an additional point. It is not difficult to see that
it is still possible to add a second higher-derivative term in the Eq.%
\eqref{eq 2.0}, but in order to preserve the original dispersion relation%
\begin{equation}
k^{2}\left( k^{2}-M^{2}\right) A_{\mu }\left( k\right) =0,
\end{equation}%
when the generalized condition $\left( k^{2}-M^{2}\right) k^{\mu }A_{\mu
}\left( k\right) =0$ holds. Hence, the Lagrangian should be rewritten as%
\begin{equation}
\mathcal{L}=-\frac{1}{4}F_{\mu \nu }F^{\mu \nu }+\frac{1}{6M^{2}}\partial
_{\mu }F^{\mu \nu }\partial ^{\lambda }F_{\lambda \nu }+\frac{1}{6M^{2}}%
\partial _{\lambda }F^{\mu \nu }\partial ^{\lambda }F_{\mu \nu },
\label{eq 2.1}
\end{equation}%
since $\left( \partial _{\lambda }F^{\mu \nu }\right) ^{2}=2\left( \partial
_{\lambda }F^{\lambda \nu }\right) ^{2}$. Therefore, the starting point of
our analysis would be the Lagrangian density \eqref{eq 2.1}. Thus, to input
an internal symmetry group, the original field must change as $A_{\mu
}\rightarrow A_{\mu }^{a}$, where $a=1,...,\left( N^{2}-1\right) $, denotes
an index belonging to some internal symmetry group introduced into the
original theory, in our cause $SU\left(N\right) $. Assuming that $%
X=X^{a}\tau ^{a}$, where $\tau ^{a}$ are the generators of the corresponding
Lie algebra, $\left[ \tau ^{a},\tau ^{b}\right] =if^{abc}\tau ^{c}$, and
that it transforms as an adjoint representation of the symmetry group, we
rewrite the original Lagrangian density
\begin{align}
\mathcal{L}_{AAB} =& -\frac{1}{4}W^{a}{}_{\mu \nu }W^{a}{}^{\mu \nu }+\frac{1}{%
6M^{2}}\left( D_{\mu }W^{\mu \nu }\right) ^{b}\left( D^{\sigma }W_{\sigma
\nu }\right) ^{b}  \notag \\
&+\frac{1}{6M^{2}}\left( D_{\sigma }W^{\mu \nu }\right) ^{a}\left(
D^{\sigma }W_{\mu \nu }\right) ^{a}-\frac{g}{18M^{2}}f^{abc}W_{\mu \nu
}^{a}W^{b\nu \lambda }W_{\lambda \alpha }^{c}\eta ^{\alpha \mu },
\label{eq 2.3}
\end{align}
where $W^{a}_{\mu \nu }$ is an non-Abelian stress-tensor with the
following form%
\begin{equation}
W^{a}{}_{\mu \nu }=F^{a}{}_{\mu \nu }+gG^{a}{}_{\mu \nu },  \label{eq 2.4}
\end{equation}%
where $g$ is a group parameter and $G^{a}{}_{\mu \nu }=f^{abc}A_{\mu
}^{b}A_{\nu }^{c}$, we also have that the covariant derivative is $D_{\mu
}^{ac}\equiv \delta ^{ac}\partial _{\mu }+gf^{abc}A_{\mu }^{b}$, with $%
\left( \tau ^{a}\right) ^{bc}=-if^{abc}$.

Now, in order to carry out the second step of the method we should first
rewrite the Lagrangian density \eqref{eq 2.3} in its first-order form. To
accomplish that we may use the canonical momenta due to algebraic
simplification in this choice. Therefore, from the definition \eqref{eq 2.6} one may
evaluate%
\begin{equation}
3M^{2}\phi ^{a\mu }=\left( D_{\sigma }W^{\sigma \mu }\right)
^{a}-\eta ^{0\mu }\left( D_{\sigma}W^{\sigma 0}\right)
^{a}+2\left( D^{0}W^{0\mu }\right) ^{a},  \label{eq 2.9}
\end{equation}%
and, it follows%
\begin{equation}
3M^{2}\phi ^{an}=\left( D_{m}W^{m n}\right) ^{a}+3\left( D_{0}W^{0n}\right)
^{a}.
\end{equation}%
With the above results in hands we may now rewrite the Lagrangian
\eqref{eq
2.3}\ in its first-order form \footnote{%
We follow again the Ostrogradski formalism to deal with the
higher-derivative terms.} as in terms of the time-derivatives of the field
potential, one then obtain the first-order Lagrangian%
\begin{equation}
\mathcal{L}=-\phi _{k}^{a}\dot{\Gamma}_{k}^{a}+\pi _{\mu }^{a}\dot{A}^{a\mu
}-\mathcal{V}^{\left( 0\right) },  \label{eq 2.11}
\end{equation}%
whereas the potential density is%
\begin{align}
\mathcal{V}^{\left( 0\right) }&= \pi _{\mu }^{a}\Gamma ^{a\mu }-\frac{M^{2}}{2}\phi _{k}^{a}\phi _{k}^{a} +\phi _{k}^{a}\bigg( \frac{1}{3}%
\left( D^{m}W_{mk}\right) ^{a}-\partial _{k}\Gamma _{0}^{a}+gf^{abc}\left(
\Gamma _{0}^{b}A_{k}^{c}+A_{0}^{b}\Gamma _{k}^{c}\right)
+gf^{abc}A_{0}^{b}W_{0k}^{c}\bigg)  \notag \\
&-\frac{1}{6M^{2}}\left( D_{0}W^{n m}\right) ^{a}\left( D^{0}W_{n m}\right)
^{a} +\frac{1}{4}%
W^{a}{}_{km}W^{a}{}^{km} -\frac{1}{2} \left( \Gamma _{k}^{a}-\partial
_{k}A_{0}^{a}+gf^{abc}A_{0}^{b}A_{k}^{c}\right) ^{2} \notag \\
& -\frac{1}{6M^{2}}\bigg[ \left( D_{r}W^{nm}\right) ^{a}\left( D^{r}W_{n m}\right) ^{a}+\left( D_{r}W^{r 0}\right) ^{a}\left(
D^{m}W_{m0}\right) ^{a} +\frac{2}{3}\left( D_{m}W^{mn}\right) ^{a}\left( D^{r}W_{rn}\right)
^{a}\notag \\
& +2\left( D_{m}W^{0r}\right) ^{a}\left(
D^{m}W_{0r}\right) ^{a} \bigg] +\frac{g}{18M^{2}}f^{abc}\left[
3W_{0m}^{a}W_{mk}^{b}W_{k0}^{c}-W_{km}^{a}W_{mj}^{b}W_{jk}^{c}\right].
\label{eq 2.11a}
\end{align}%
From the above expression \eqref{eq 2.11} for the Lagrangian density one may
reads the initial set of sympletic variables%
\begin{equation}
\xi _{\alpha }^{\left( 0\right) }=\left\{ A_{k}^{a},\pi
_{k}^{a},A_{0}^{a},\pi _{0}^{a},\Gamma _{k}^{a},\phi _{k}^{a},\Gamma
_{0}^{a}\right\} ,  \label{eq 2.12}
\end{equation}%
moreover, from \eqref{eq 2.11} we can identify the non-null canonical
one-form%
\begin{equation}
^{\Gamma }a_{i}^{\left( 0\right) }=-\phi _{i}^{a},\quad ^{A}a_{i}^{\left(
0\right) }=-\pi _{i}^{a},\quad ^{A_{0}}a^{\left( 0\right) }=\pi _{0}^{a}.
\label{eq 2.13}
\end{equation}%
From these results, we can compute the elements of the sympletic matrix,
leading to the corresponding two-form matrix%
\begin{equation}
^{\left( 0\right) }f_{ab}\left( x,y\right) =\left[
\begin{array}{cc}
\mathbf{A}_{ij}^{ab} ~ & ~\mathbf{0}_{\mathbf{4\times 3}} \\
\mathbf{0}_{\mathbf{3\times 4}} ~&~ \mathbf{B}_{ij}^{ab}
\end{array}%
\right] \delta \left( x,y\right) ,  \label{eq 2.16}
\end{equation}%
with $\mathbf{A}_{ab}\left( x,y\right) $ and $\mathbf{B}_{ab}\left(
x,y\right) $ having the same expression to the $\mathbf{A}\left( x,y\right) $
and $\mathbf{B}\left( x,y\right) $, Eq.\eqref{eq 0.12}, with an additional
non-Abelian index $\delta ^{ab}$. The matrix $^{\left( 0\right)
}f_{ab}\left( x,y\right) $ is obviously singular. The eigenvector with zero
eigenvalue is%
\begin{equation}
\nu _{\alpha }=\left( \mathbf{0},\mathbf{0},0,0,\mathbf{0},\mathbf{0},\nu
_{7}^{a}\right) ,  \label{eq 2.17}
\end{equation}%
where $\nu _{7}$ is an arbitrary function and associated with $\Gamma
_{0}^{a}$. Hence, from the eigenvector \eqref{eq 2.17} we can calculate the
consistence condition%
\begin{equation}
\int dx\nu _{7}^{a}\left( x\right) \left[ \pi _{0}^{a}+D_{k}^{ab}\phi
_{k}^{b}\right] \left( x\right) =0,
\end{equation}%
and, since $\nu _{7}^{a}$ is an arbitrary function, we obtain the constraint%
\begin{equation}
\chi ^{a}\left( x\right) \equiv \pi _{0}^{a}+D_{k}^{ab}\phi _{k}^{b}=0.
\label{eq 2.18}
\end{equation}%
Following the methodology, we should now introduce this constraint back into
the Lagrangian by means of a Lagrange multiplier $\lambda $, one then gets%
\begin{equation}
\mathcal{L}=-\phi _{k}^{a}\dot{\Gamma}_{k}^{a}+\pi _{\mu }^{a}\dot{A}^{a\mu
}+\dot{\lambda}^{a}\left( \pi _{0}^{a}+D_{k}^{ab}\phi _{k}^{b}\right) -%
\mathcal{V}^{\left( 1\right) },  \label{eq 2.19}
\end{equation}%
where the first-iterated potential density is $\mathcal{V}^{\left( 1\right)
}=\left. \mathcal{V}^{\left( 0\right) }\right\vert _{\chi =0}$. From the
above Lagrangian we may notice that the field $\Gamma _{0}$ naturally
disappears when the constraint is taken as a strong relation. Now, in the
first-iterate case, the sympletic variables are $\xi _{\alpha }^{\left(
1\right) }=\left\{ A_{k}^{a},\pi _{k}^{a},A_{0}^{a},\pi _{0}^{a},\Gamma
_{k}^{a},\phi _{k}^{a},\lambda ^{a}\right\} $, and we can read the following
one-form%
\begin{gather}
^{\Gamma }a_{i}^{\left( 1\right) } =-\phi _{i}^{a}, \quad ^{A}a_{i}^{\left(
1\right) }=-\pi _{i}^{a},\quad  ^{A_{0}}a^{\left( 1\right) }=\pi _{0}^{a}, \quad
  ^{\lambda }a^{\left( 1\right) }=\pi _{0}^{a}+D_{k}^{ab}\phi
_{k}^{b}.  \label{eq 2.21}
\end{gather}%
By evaluating the corresponding matrix elements, the sympletic two-form
matrix reads%
\begin{equation}
^{\left( 1\right) }f\left( x,y\right) =\left[
\begin{array}{cc}
\mathbf{A}^{ab}_{ij}  ~&~ \mathbf{D}^{ab}_{j}  \\
-(\mathbf{D}^{T})^{ba}_{i} ~&~ \mathbf{C}^{ab}_{ij}
\end{array}%
\right] \delta \left( x,y\right),  \label{eq 2.22}
\end{equation}%
with%
\begin{align}
\mathbf{C}^{ab}_{ij} =\left[
\begin{array}{ccc}
0 & \delta ^{ab}\delta _{ij} & 0 \\
-\delta ^{ab}\delta _{ij} & 0 & \left( D_{y}\right) _{i}^{ba} \\
0 & \left( D_{x}\right) _{i}^{ab} & 0%
\end{array}%
\right]  ,  \quad \mathbf{D}^{ab}_{i} =%
\left[
\begin{array}{ccc}
0 & 0 & -gf^{abc}\phi _{i}^{c} \\
0 & 0 & 0 \\
0 & 0 & 0 \\
0 & 0 & \delta ^{ab}%
\end{array}%
\right]  .  \label{eq 2.24}
\end{align}%
Again, we see that the first-iterated sympletic is singular. In the next
step we should determine its eigenvector with zero eigenvalue. From that it
follows%
\begin{equation}
\overline{\nu }_{\alpha }=\left( \mathbf{0},\left( \overline{\nu }%
_{2}\right) _{k}^{a},\left( \overline{\nu }_{3}\right) ^{a},0,\left(
\overline{\nu }_{5}\right) _{k}^{a},\mathbf{0},\left( \overline{\nu }%
_{7}\right) ^{a}\right) ,  \label{eq 2.23}
\end{equation}%
Therefore, from the eigenvector $\overline{\nu }_{\alpha }$ \eqref{eq 2.23}
we can evaluate the consistence condition and gets%
\begin{align}
\int dx\nu _{7}^{b}\left( x\right) \bigg[ \left( D_{k}\right) ^{bc}\left(
\pi ^{ck}-gf^{cde}\phi ^{dk}A_{0}^{e}\right) +
+gf^{bdc}\phi _{k}^{d}W_{0k}^{c}\bigg] \left( x\right)  = 0 ,
\end{align}%
and, since $\nu _{7}^{a}$ is an arbitrary function, we obtain the constraint%
\begin{equation}
\bar{\chi}^{b}\equiv \left( D_{k}\right) ^{bc}\left( \pi
^{ck}-gf^{cde}\phi ^{dk}A_{0}^{e}\right) +gf^{bdc}\phi _{k}^{d}W_{0k}^{c}=0,
\label{eq 2.29}
\end{equation}%
which is nothing more than the non-Abelian version of the Gauss's law.
Proceeding, we should introduce it back to the Lagrangian as a strong
relation, thus the second-iterated Lagrangian reads%
\begin{align}
\mathcal{L}=& \dot{\eta}^{b}\left( \left( D_{k}\right) ^{bc}\left( \pi
^{ck}-gf^{cde}\phi ^{dk}A_{0}^{e}\right)+gf^{bdc}\phi _{k}^{d}W_{0k}^{c}\right) \notag \\
& -\phi _{k}\dot{\Gamma}_{k}+\pi _{\mu }\dot{A}^{\mu }+\dot{%
\lambda}\left( \pi _{0}^{a}+\left( D_{k}\phi _{k}\right) ^{a}\right)   -\mathcal{V}^{\left( 2\right) },  \label{eq 2.30}
\end{align}%
whereas the second-iterated potential density is given by $\mathcal{V}^{\left( 2\right) }= \left. \mathcal{V}^{\left( 1\right)
}\right\vert _{\bar{\chi}=0} $ with
\begin{align}
\mathcal{V}^{\left( 2\right) }&= -\pi _{k}^{a}\Gamma _{k}^{a}
-\frac{M^{2}}{2}\phi _{k}^{a}\phi _{k}^{a} +\phi _{k}^{a}\left( \frac{1}{3}%
\left( D^{m}W_{mk}\right) ^{a}+gf^{abc}A_{0}^{b}\Gamma
_{k}^{c}+gf^{abc}A_{0}^{b}W_{0k}^{c}\right)  \notag \\
&+\frac{1}{4}W^{a}{}_{km}W^{a}{}^{km} -\frac{1}{2}\left( \Gamma _{k}^{a}-\partial
_{k}A_{0}^{a}+gf^{abc}A_{0}^{b}A_{k}^{c}\right) ^{2}-\frac{1}{6M^{2}}\left( D_{0}W^{nm}\right) ^{a}\left( D^{0}W_{nm}\right)
^{a} \notag \\
& -\frac{1}{6M^{2}}\bigg[\left( D_{r}W^{n m}\right) ^{a}\left( D^{r}W_{m m}\right) ^{a}+ \left( D_{r}W^{r0}\right) ^{a}\left(
D^{m}W_{m0}\right) ^{a}+2\left( D_{m}W^{0n}\right) ^{a}\left(
D^{m}W_{0n}\right) ^{a}\notag \\
& +\frac{2}{3}\left( D_{m}W^{m n}\right) ^{a}\left( D^{r}W_{rn}\right)
^{a}\bigg] +\frac{g}{18M^{2}}f^{abc}\left[
3W_{0m}^{a}W_{mk}^{b}W_{k0}^{c}-W_{km}^{a}W_{mj}^{b}W_{jk}^{c}\right]. \label{eq 2.31}
\end{align}
From the above second-iterated Lagrangian we read the following vectors
\begin{gather}
^{\Gamma }a_{i}^{\left( 2\right) } =-\phi _{i},\quad ^{A}a_{i}^{\left(
2\right) }=-\pi _{i},\quad ^{A_{0}}a^{\left( 2\right) }=\pi _{0}, \quad
^{\lambda }a^{\left( 2\right) }=\pi _{0}^{a}+\left( D_{k}\phi _{k}\right)
^{a},  \notag \\
^{\eta }a^{\left( 2\right) } =\left( D_{k}\right) ^{bc}\left( \pi
^{ck}-gf^{cde}\phi ^{dk}A_{0}^{e}\right) +gf^{bdc}\phi _{k}^{d}W_{0k}^{c},
\end{gather}%
these results lead to the corresponding two-form matrix for $\xi _{\alpha
}^{\left( 2\right) }=\left\{ A_{k}^{a},\pi _{k}^{a},A_{0}^{a},\pi
_{0}^{a},\Gamma _{k}^{a},\phi _{k}^{a},\lambda ,\eta \right\} $
\begin{equation}
^{\left( 2\right) }f\left( x,y\right) =\left[
\begin{array}{cc}
\mathbf{A}^{ab}_{ij} ~ & ~ \mathbf{E}^{ab}_{j,x}   \\
-(\mathbf{E}^{T})_{i,y}^{ba} ~ & ~ \mathbf{F}^{ab}_{ij}
\end{array}%
\right] \delta\left( x,y\right),
\end{equation}%
with
\begin{align}
\mathbf{E}^{ab}_{i,x}& =\left[
\begin{array}{cccc}
0 & 0 & -gf^{abc}\phi _{i}^{c}  & ^{\left( 2\right)
\left( A,\eta \right) }f^{ab}  \\
0 & 0 & 0 & \left( D_{x}^{i}\right) ^{ba}  \\
0 & 0 & 0 & ^{\left( 2\right) \left( A_{0},\eta \right) }f^{ab}  \\
0 & 0 & \delta ^{ab}  & 0%
\end{array}%
\right] , \\
\mathbf{F}^{ab}_{ij} & =\left[
\begin{array}{cccc}
0 & \delta ^{ab}\delta _{ij}  & 0 & gf^{adb}\phi
_{i}^{d}  \\
-\delta ^{ab}\delta _{ij}  & 0 & \left( D^{y}\right)
_{i}^{ba} & ^{\left( 2\right) \left( \phi ,\eta
\right) }f^{ab}  \\
0 & -\left( D_{i}^{x}\right) ^{ab}  & 0 & 0 \\
-gf^{adb}\phi _{i}^{d} & ^{\left( 2\right) \left(
\eta ,\phi \right) }f^{ab}   & 0 & 0%
\end{array}%
\right] .
\end{align}%
We have again obtained a singular matrix, now the second-iterated one $%
^{\left( 2\right) }f\left( x,y\right) $. Next, we determine the zero-mode
vectors, which now consist in a set of two vectors
\begin{equation}
\tilde{\nu}_{\alpha }=\left( \left( \overline{\overline{\nu }}_{1}\right)
_{k}^{a},\mathbf{0},0,\left( \overline{\overline{\nu }}_{4}\right) ^{a},%
\mathbf{0},\left( \overline{\overline{\nu }}_{6}\right) _{k}^{a},0,\left(
\overline{\overline{\nu }}_{8}\right) ^{a}\right)
\end{equation}%
and%
\begin{equation}
\overline{\overline{\nu }}_{\alpha }=\left( \mathbf{0},\left( \overline{%
\overline{\nu }}_{2}\right) _{k}^{a},\left( \overline{\overline{\nu }}%
_{3}\right) ^{a},0,\left( \overline{\overline{\nu }}_{5}\right) _{k}^{a},%
\mathbf{0},\left( \overline{\overline{\nu }}_{7}\right) ^{a},0\right) .
\end{equation}%
However, the vector $\overline{\overline{\nu }}_{\alpha }$ generates the
constraint $\bar{\Omega}^{b}\left( x\right) \equiv \left( D_{k}\right)
^{bc}\left( \pi ^{ck}-gf^{cde}\phi ^{dk}A_{0}^{e}\right) +gf^{bdc}\phi
_{k}^{d}W_{0k}^{c}=0$. Therefore, it is the vector $\tilde{\nu}_{\alpha }$
only of our interest; but this zero-mode does not generate any new
constraints and, consequently, the sympletic matrix remains singular.
Therefore, there are gauge degrees of freedom to be fixed. A suitable choice
here is the generalized Coulomb gauge $\left( 1+M^{-2}\square \right)
\partial ^{k}A_{k}^{a} =0$, $\Gamma _{0}^{a}=0$ and $A_{0}^{a}=0$.
The same set of constraints obtained here was previously found in \cite{13}
through an analysis \textit{\`{a} la} Dirac, where it was also discussed the
generator of the gauge symmetry as well. This shows again an accordance
between the Dirac and Faddeev-Jackiw methods.

\subsection{Transition-amplitude via BFV}

Instead of evaluating the inverse of the third-iterated two-form matrix, and
then determine the generalized brackets between the dynamical fields, and
proceed to the quantization by the correspondence principle, we shall rather
work in a path-integral framework. Therefore, we shall now compute the transition-amplitude through the BFV
formalism \cite{22}, because, though there is a proposal relating the
path-integral to the FJ-method \cite{8}, it is not clear that the method
works and it is consistent to a gauge theory. We have that the
transition-amplitude in our case is written%
\begin{align}
\mathcal{Z} =& \int DA_{k}^{a}D\pi _{k}^{a}D\Gamma _{m}^{b}D\phi _{m}^{b}D\lambda
^{c}Db^{c}D\bar{c}^{d}Dc^{d}D\bar{P}^{e}DP^{e}  \label{eq 3.1} \\
&\times \exp \bigg[ i\int d^{4}x\bigg\{ \pi _{k}^{a}\dot{A}^{ak}+\phi
_{k}^{a}\dot{\Gamma}^{ak}+\dot{c}^{a}\bar{P}^{a}+\left( \partial _{0}\bar{c}%
\right) ^{d}P^{d}+\dot{\lambda}^{a}b^{a}-\mathcal{V}^{\left( 3\right)
}\bigg\} +i\int dx_{0}\left\{ \Psi ,Q_{BRST}\right\} \bigg]  \notag
\end{align}%
where $\mathcal{V}^{\left( 3\right) }$ is recognized as being the canonical
Hamiltonian in the first-order approach, and it is given by the Eq.\eqref{eq
2.31}
\begin{equation}
\mathcal{V}^{\left( 3\right) }=\left. \mathcal{V}^{\left( 2\right)
}\right\vert _{\Omega =0},
\end{equation}%
also, $\Omega $ consists in the generalized Coulomb gauge
\begin{equation}
\left( 1+M^{-2}\square \right)  \partial ^{k}A_{k}^{a} =0,
\quad \Gamma _{0}^{a}=0, \quad A_{0}^{a}=0.
\end{equation}%
Furthermore, $\left( c_{a},\bar{P}_{a}\right) $ and $\left( \bar{c}%
_{a},P_{a}\right) $, are the pairs of ghost fields and their respective
momenta, while $\left( \lambda _{a},b_{a}\right) $ is a Lagrange multiplier
and its momentum, all satisfying the following Berezin brackets:%
\begin{gather}
\left\{ \bar{c}_{a}\left( z\right) ,P_{b}\left( w\right) \right\}
_{B}=\delta _{ab}\delta \left( z,w\right) , \quad
\left\{ \bar{P}_{a}\left( z\right) ,c_{b}\left( w\right) \right\} _{B}=-\delta _{ab}\delta \left(
z,w\right) ,\notag \\
\left\{ \lambda _{a}\left( z\right) ,b_{b}\left( w\right)
\right\} _{B}=\delta _{ab}\delta \left( z,w\right) .
\end{gather}%
In the expression \eqref{eq 3.1} it remains to define two quantities. The
first is the BRST charge, which with the full set of constraints, is written
\begin{align}
Q_{BRST}=\int d^{3}x\bigg[ c^{b}\left( \left( D_{k}\pi ^{k}\right)
^{b}+gf^{bdc}\phi _{k}^{d}\Gamma _{k}^{c}\right) -iP^{a}b^{a}+\frac{1}{2}%
\overline{P}^{a}f^{abc}c^{b}c^{c}\bigg],
\end{align}%
whereas we have that the gauge-fixing function $\Psi $, in the generalized Coulomb gauge,
reads
\begin{align}
\Psi =\int d^{3}z\bigg[ \frac{i\xi }{2}b^{a}\bar{c}^{a}+i\bar{c}^{a}\left(
1+M^{-2}\square \right) \partial ^{k}A_{k}^{a}-\lambda ^{a}\left(
1+M^{-2}\square \right) ^{-1}\bar{P}^{a}\bigg] .
\end{align}%
From the above expressions, it is not complicated to evaluate
\begin{align}
\left\{ \Psi ,Q_{BRST}\right\} =&\int d^{3}z\bigg\{\frac{\xi }{2}%
b^{a}b^{a}+i\partial ^{k}\bar{c}^{e}\left( 1+M^{-2}\square \right) \left(
D_{k}c\right) ^{e}+b^{a}\left( 1+M^{-2}\square \right) \partial ^{k}A_{k}^{a}\notag \\
&+i\left( 1+M^{-2}\square \right) ^{-1}\overline{P}^{a}P^{a}+f^{abc}\left(
1+M^{-2}\square \right) ^{-1}\overline{P}^{a}\lambda ^{b}c^{c}  \notag \\
&+\left( 1+M^{-2}\square \right) ^{-1}\lambda ^{a}\left( \left( D_{k}\pi
^{k}\right) ^{b}+gf^{bdc}\phi _{k}^{d}\Gamma _{k}^{c}\right) \bigg\}.
\label{eq 3.3}
\end{align}%
Thus, substituting the result \eqref{eq 3.3} into the transition-amplitude
expression \eqref{eq 3.1}, and performing the variables integration and after some algebraic manipulation,
one finds
\begin{align}
\mathcal{Z} =&\int DA_{\mu }^{a}D\bar{c}^{d}Dc^{d} \notag \\
&\times \exp \bigg[i\int d^{4}z\bigg\{%
\mathcal{L}_{AAB} + i\partial _{\mu }\bar{c}^{e}\left( 1+M^{-2}\square \right) \left( D^{\mu
}c\right) ^{e}  -\frac{1}{2\xi }\left[ \left( 1+M^{-2}\square \right)
\partial ^{\mu }A_{\mu }^{a} \right] ^{2}\bigg\}\bigg].
\label{eq 3.6}
\end{align}
Hence, from the BFV formalism we have obtained directly the desirable
covariant expression for the transition-amplitude. Furthermore, we see that
the ghosts fields are coupled from the gauge fields, and matter fields may
also be included.


\section{Concluding Remarks}

\label{sec:6}

In this paper we have presented a canonical study of higher-derivative theories, the Podolsky's
electrodynamics and its non-Abelian extension, the Alekseev-Arbuzov-Baikov's
effective Lagrangian, in the point of view of the sympletic Faddeev-Jackiw
approach. Although the Dirac's method remains as the standard method to deal
with constrained systems, it has been recognized that some calculation is
unnecessarily cumbersome there, and then it is exactly there where the
FJ method shows to be an economical and rich framework for
first-order Lagrangian functions, obviating mainly unnecessary calculations.

At the beginning we have reviewed briefly the main aspects of the sympletic
FJ method. Subsequently, we applied the method on studying the generalized
electrodynamics. The full set of the known constraints \cite{12} was
obtained, and afterwards it was showed that the third zero-mode vector does
not generate any new constraint, however the sympletic matrix remained
singular, an imprint characteristic of gauge theories. Therefore, the gauge
had to be fixed, to attain that we had chosen to work with the generalized
Coulomb gauge. From all that we were able to obtain a nonsingular sympletic
matrix, and by evaluating the inverse of such a matrix, we obtained the
generalized brackets between the dynamical fields, in accordance with the
previous results of Dirac's approach in \cite{12}. Moreover, next we
introduced the AAB's effective Lagrangian. The lines in
studying this non-Abelian theory followed those presented to the generalized
electrodynamics. Again, the known full set of constraints was obtained, in
accordance with the result from the Dirac's approach \cite{13}. As it
happens in gauge theories, the third zero-mode vector does not generate any
new constraint and the sympletic matrix remained singular. However, instead
of using the usual prescription and introduce new constraints, in order to
fix the gauge degrees-of-freedom, we had chosen to quantize the theory via
path-integral methods. Although it is known a proposal of FJ method in the
path-integral framework, its content it is not clear to gauge theories.
Therefore, we followed the well-known BFV-method to construct the transition-amplitude.

It was successfully showed here that the sympletic approach of
Faddeev-Jackiw works perfectly also to higher-derivative theories, and that
all the obtained results were in accordance with previous ones when the
Dirac's methodology was applied. This is somehow in contrast with the
assertion \cite{5} that the FJ method produces constraints that do not exist
in the Dirac's theory. Furthermore, this emphasizes that in fact the FJ
method poses as a good candidate of framework where deeper analysis may be
performed, especially in more intriguing theories, such as General
Relativity and renormalizable higher-derivative proposals of a quantum
theory of Gravity, in different dimensionality, where the constraint
analysis is not always easy to accomplish and clear within the Dirac's
methodology. These issues and others will be further elaborated,
investigated and reported elsewhere.

\subsection*{Acknowledgments}

RB thanks FAPESP for full support, BMP thanks CNPq and CAPES for partial
support.



\end{document}